\documentclass[prl,preprint,aps]{revtex4}
\usepackage{amsmath,bm}
\usepackage{graphicx}

\begin{document}

\title{Topological thermal instability and length of proteins}

\author{Raffaella Burioni}
\affiliation{Dipartimento di Fisica and INFM, Universit\`a di Parma, 
Parco Area delle Scienze 7A, 43100 Parma, Italy}
\author{Davide Cassi}
\affiliation{Dipartimento di Fisica and INFM, Universit\`a di Parma, 
Parco Area delle Scienze 7A, 43100 Parma, Italy}
\author{Fabio Cecconi}
\affiliation{Dipartimento di Fisica and INFM (UdR and SMC) , Universit\`a
di Roma La Sapienza,
P.le A. Moro 2, 00185 Roma, Italy}
\author{Angelo Vulpiani}
\affiliation{Dipartimento di Fisica and INFM (UdR and SMC) , Universit\`a
di Roma La Sapienza, P.le A. Moro 2, 00185 Roma, Italy}

\begin{abstract}
We present an analysis of the effects of
global topology on the structural stability of folded proteins in thermal
equilibrium with a heat bath. For a large class of single domain proteins, we
computed the harmonic spectrum within the Gaussian Network Model (GNM) and
determined the spectral dimension, a parameter describing the low frequency
behaviour of the density of modes. 
We find a surprisingly strong correlation
between the spectral dimension and the number of amino acids of the protein. 
Considering
that larger spectral dimension value relate to more topologically compact
folded state, our results indicate that for a given temperature and length of
the protein, the folded structure corresponds to the less compact folding
compatible with thermodynamic stability.
\vskip 1 truecm

\end{abstract}
\maketitle

\section{INTRODUCTION}
The role of geometry has recently been considered as a factor of primary 
importance for the study of several physical properties of 
proteins 
and other biological macromolecules. 
In particular, 
since the topology of folded states is known to influence the folding 
properties of the protein
\cite{Go,Plaxco,Makarov,Baker,Riddle,Clementi,Thirumalai,Cecconi}, a great 
deal of work has been devoted to the study of those theoretical 
aspects which describe the
networks of links between amino acids in folded proteins. 
\cite{Kaba,Park,Vendru}. 
Furthermore, relevant features of protein conformations seem to follow 
the geometrical principles of the optimal packing 
problem~\cite{Maritan,Banavar} and mathematical concepts from graph 
theory have been interestingly applied to identify flexible and rigid 
regions of folded states~\cite{Thorpe}.

Starting from the primary, linear structure (the
sequence of amino acids), a protein evolves during the folding process 
until it reaches a final state (native state) whose geometrical shape is 
crucial to the function of the protein itself. However, the problem of the 
geometrical arrangement of proteins in their native states cannot be 
regarded as purely static issue. Indeed, a massive accumulation of 
experimental data collected from X-ray, NMR and neutron spectroscopy,
has revealed that protein native states are rather 
dynamic structures where amino acids constantly move around their
equilibrium positions. 
This motion, crucially involved in protein 
functions \cite{Frauen,Frauen2}, is usually examined and 
investigated through normal modes analysis (NMA) \cite{NMA}
or essential dynamics \cite{Essential}.
However, the study of collective motions of 
large scale proteins is generally difficult due to limited access to realistic 
all-atoms 
NMA~\cite{Levitt} and simplified or approximate approaches are usually 
welcome. 
Tirion \cite{Tirion} first
proposed the possibility of replacing, in protein normal mode computations, 
complicated empirical potentials by Hookian pairwise interactions depending 
on a single parameter. 
This approach stems from the observation that low-frequency dynamics,
which are mainly associated with protein-domain motion, are generally
insensitive to the finer details of atomic interactions.
Much of the subsequent literature \cite{Hinsen,Bahar,Atilgan,
Haliloglu,Cecco2,Miche2} has 
confirmed
the success of simple harmonic models in the study of slow vibrational 
dynamics of large biological macromolecules, and they have become 
a viable alternative to heavy and time-consuming all-atoms NMA. 
This success result from the striking agreement of predictions with experiments, the 
presence of few adjustable parameters and  the fast and easy numerical 
implementation on computers and fast result production.
For these reasons harmonic models are also utilized for the
systematic analysis of large data sets of proteins.

The topological stability of macromolecules is far from being
a pure mechanical problem as it closely involves thermodynamics. Indeed,
the relevant thermodynamic potential that must be minimized in order to find 
the stable configuration is not energy, 
but free energy. This is due to the interaction with
the environment (schematized as a thermal bath) which is generally not  
negligible, especially for biological macromolecules that have a stable 
phase in a solvent. In particular, water is a very efficient medium 
for the transfer of thermal energy at microscopic scales 
(i.e. oscillations and molecular rotations).

With these considerations in mind, in this work we apply NMA to investigate 
the influence of the global native state topology on   
the thermal stability of proteins.

Vibrational thermal instability is a well-known topic of study in solid state physics. 
Since the initial classical analysis of Peierls \cite{Peierls}, 
it has been recognized 
that equilibrium with a thermal bath can dramatically influence the 
possible topological arrangement of large geometrical structures. 
Up to now, the most striking consequence of Peierls' instability has concerned  low-dimensional crystals: for one and two-dimensional 
lattices the mean square
displacement of a single atom at finite temperature diverges in the
thermodynamic limit, i.e. with an increasing number of atoms. 
When the displacement exceeds the order of magnitude of the lattice 
spacing, the topological arrangement of the lattice is unstable and the 
crystal becomes a liquid. For real structures, formed by a finite 
number of units and far from the
thermodynamic limit, the divergence sets a maximal stability size, which is 
negligible for one-dimensional lattices and typically
mesoscopic for two-dimensional lattices.
 
However, thermal instability is present not only in crystals but also 
in structurally inhomogeneous systems, such as glasses, 
fractals, polymers and non crystalline structures. Here, 
the problem is much more complex. 
Generalizing the Peierls approach to mesoscopic disordered structures, 
we are able to apply this kind of argument to the thermal stability of 
macromolecules. In this article we describe how this can be done in the 
case of proteins; we  predict the existence of a critical stability 
size depending on a global topological parameter (the spectral dimension) 
and compare our predictions with experimental data.

\section{THEORY}

In a recent paper \cite{BCFV} generalizing the Peierls' result, we
showed that a thermodynamic instability also appears in inhomogeneous 
structures and is determined by the spectral dimension $\bar {d}$. 
The parameter $\bar{d}$ \cite{Orbach} is defined
according to the asymptotic behaviour of the density of harmonic 
oscillations at low frequencies.
More precisely, using $g(\omega)$ to denote the density of modes with 
frequency $\omega$, then
\begin{equation}
g(\omega) \sim \omega^{\bar{d}-1}
\label{eq:dos}
\end{equation}
for $\omega\to 0$.
The spectral dimension is the most natural 
extension of the usual Euclidean dimension $d$ to disordered 
structures as far as dynamical processes are concerned. 
It coincides with $d$ in the case of lattices, but in general, it can 
assume non-integer values between $1$ and $3$. 
The spectral dimension represents a useful measure of the
effective connectedness of geometrical structures at large scales, because 
large values of $\bar{d}$ correspond to high topological 
connectedness. 
Moreover, it characterizes not only
harmonic oscillations, but it also relates to diffusion, 
phase transitions and electrical conductivity, allowing a variety of
both experimental and numerical methods for its determination
\cite{Burioni,Saviot}.
The relevance of $\bar{d}$ in connection with the anomalous density of 
vibrational modes in proteins has also been considered in 
refs.\cite{Avraham,Karplus}.

In the case of thermal instability, we demonstrated that, 
for $\bar {d} \ge 2$, the mean
square displacement $\langle r^2 \rangle$ of a structural unit (being an
atom, a molecule or a supra-molecular structure according to the studied case)
of a system composed of $N$ elements, diverges in the limit $N\to\infty$.
Using $T$ to denote the temperature of the heat bath, with $k_B$ the
Boltzmann constant, and with $\gamma$ the interaction energy scale,
the divergence is given by the asymptotic law:
\begin{equation}
\langle r^2 \rangle \sim \frac{k_B T}{\gamma}  N^{2/\bar{d} - 1}
\label{PL}
\end{equation}
when $\bar{d} < 2$. When $\bar {d} = 2$, the
mean square displacement diverges logarithmically, $\langle r^2 \rangle \sim
k_B T/\gamma \ln(N)$, as in the case of the Peierls' result for a two dimensional
crystal. Notice that the divergence in $\langle r^2 \rangle$
is only determined by $\bar {d}$.
Now, at any
given temperature $T$, there will exist a threshold value $N(T)$ beyond
which $\langle r^2 \rangle^{1/2}$ exceeds the typical spacing
between the nearest neighbors, making the solid structure unstable. 
Therefore at large enough values of $N$, the solid will experience a structural 
reorganization
which can lead either to a homogeneous liquid phase at sufficiently high 
temperatures
or to a disordered 3-dimensional solid, which is homogeneous on a large scale
and inhomogeneous at a small scale.
In general,
the threshold values of $N$ are very small with respect to the typical
order of magnitude of macroscopic systems, being rather comparable to the size
of large complex macromolecules such as biopolymers. 

This poses an intriguing question concerning proteins. 
Indeed, to exploit their biological function
proteins must keep a specified geometrical and topological 
arrangement and cannot
afford any, even partial, large scale geometrical fluctuations such as it happen,
to swollen polymeric chains in a good solvent \cite{Degennes}. 
This makes thermodynamical stability 
crucial and suggests a possible correlation between the spectral 
dimension and the length of protein chains. 

Vibrational stability in proteins has been analysed with 
the Gaussian network model (GNM), proposed by Bahar et al. \cite{GNM}
and widely applied because it yields results in agreement with principal X-ray 
spectroscopy experiments. 
This approach generally considers proteins as elastic networks, whose nodes 
are the positions of the alpha-carbons (C$_{\alpha}$) in the
native structure and the interactions between nodes are assimilated to
harmonic
springs. The only information required to implement the method is
the knowledge of the native structure, and two parameters 
are introduced, the spring constant and the interaction cutoff, which, 
however turn out to be related whenever the model is applyed 
to fit experimental data.
The GNM can be defined by the quadratic Hamiltonian 
\begin{equation}
H = \sum_i^{N} \frac{{\bf p}_i^2}{2 m} +
\frac{\gamma}{2}\sum_{ij} \Delta_{ij}(\delta{\bf r}_i - \delta{\bf r}_j)^2
\label{eq:GNM}
\end{equation}
where the first term is the kinetic energy of the system,  
$\gamma$ being the strength of the springs that are assumed 
homogeneous, ${\bf R}_i$ and $\delta {\bf r}_i$ indicating 
the equilibrium position and the displacement with respect to ${\bf R}_i$
of the $i$-th  C$_{\alpha}$ atoms. The model is eventually defined by the 
contact
matrix ${\Delta}$ with
entries: $\Delta_{ij}=1$ if the distance $|{\bf R}_i - {\bf R}_j|$
between two C$_{\alpha}$'s, in the native conformation,
is below the cutoff $R_0$, while is $0$ otherwise. 

The harmonic spectrum for each structure is given by the set of 
eigenvalues $\{\omega_1,...,\omega_N\}$ of the
Kirchhoff matrix (or valency-adjacency matrix)
$\Gamma_{ij} = - \Delta_{ij} + \delta_{ij} \sum_{l\neq i} \Delta_{il}$,

Notice that
the first eigenvalue $\omega_1$ vanishes and corresponds to the constant 
eigenvector related to the trivial uniform translation.

The comparison between experimental data and GNM results is obtained via the
X-ray crystallographic $B$-factors, measuring the mean square
fluctuation of C$_\alpha$ atoms around their native positions

$$
B_i(T) = \frac{8\pi^2}{3}\langle \delta{\bf r}_i \cdot \delta{\bf r}_i \rangle
$$

with $\langle \cdot \rangle$ indicating the thermal average.
In the GNM approximation, this average is easily carried out, because
amounts to a Gaussian integration, and B-factors can be expressed in
terms of the diagonal part of the inverse of the matrix $\Gamma$~\cite{GNM}:  

$$
\langle \delta{\bf r}_i \cdot \delta{\bf r}_j \rangle = \frac{3 k_B T}{\gamma}
[\Gamma^{-1}]_{ij}
$$

The knowledge of the eigenvectors and eigenmodes of matrix $\Gamma$ 
allows to compute the GNM B-factors also through formula
$$
B_i(T) = \frac{8\pi^2 k_B T}{\gamma} \sum_{k} \frac{|u_i(k)|^2}{\omega_k^2}
$$
where $i$ is the residue index, the sum runs over all non-zero 
frequencies $\omega_i$ and 
$u_i(k)$ indicates the $i$-th component of the $k$-th eigenmode.

The comparison with crystallographic data is crucial for 
setting the correct values of the parameters $R_0$ end $\gamma$.
(see. Methods and Results).

\section{METHODS}

We present a GNM harmonic analysis performed over the dataset 
of protein native structures with different sizes downloaded from the
Brookheaven Protein Data Bank.
The purpose of the analysis is basically to
investigate whether there exists a correlation between the spectral
dimension of native structures and the length of natural occurring proteins and,
if so, to verify whether the correlation can be 
explained in terms of the above mentioned stability criterion
determined by equation~(\ref{PL}).

Our representative statistical sample, 
listed in Tables \ref{tab7} and \ref{tab6}, was selected
according to the following criteria.
First, we only considered proteins with a stable large 
scale geometry. This excludes multiple domains proteins, where domains can 
undergo relative motion giving rise to larger geometrical fluctuations. 
Moreover, we considered only proteins not binded to fragments of
DNA, RNA or other substrates because such structures cannot be described 
with sufficient accuracy in terms of simple
harmonic model with only two effective parameters. 
Finally, we choose proteins covering uniformly a wide length interval 
ranging from $100$ to $3600$ to test our prediction.

The diagonalization of the Kirchhoff matrix $\Gamma$ to obtain 
its eigenvalues $\{\omega_1^2,...,\omega_n^2\}$ and eigenvectors has been 
performed with the standard numerical packages \cite{Recipes}. 

\begin{figure}
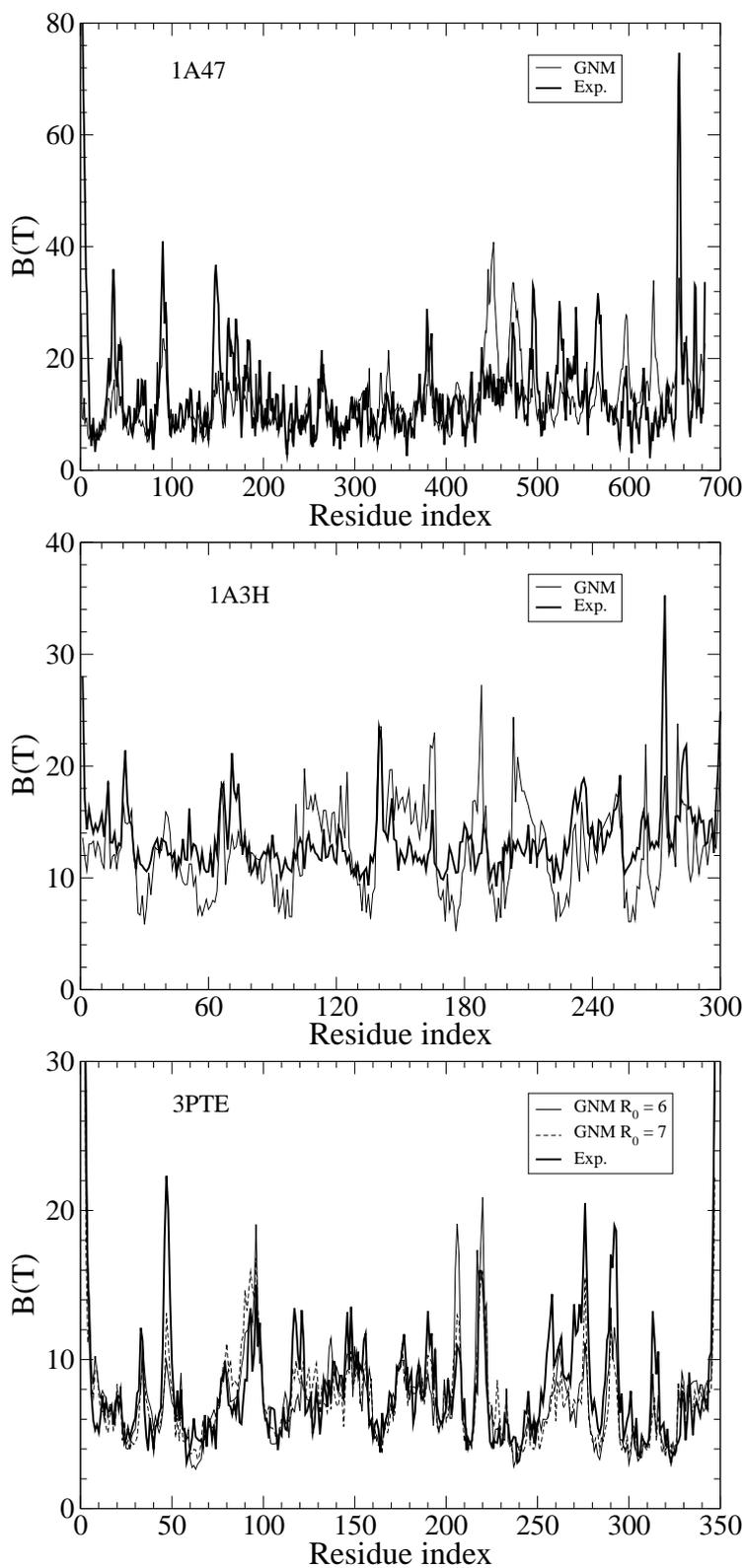

\includegraphics[clip=true,width=0.6\columnwidth,keepaspectratio]
{bfc1A47_c6.eps}
\includegraphics[clip=true,width=0.6\columnwidth,keepaspectratio]
{bfc1A3H_c6.eps}
\includegraphics[clip=true,width=0.6\columnwidth,keepaspectratio]
{bfc3PTE.eps}
\caption{\label{fig:bfact} Comparison between
experimental B-factors and mean square fluctuations of C$_\alpha$
by GNM, for the structures
1A47 (lowest correlation) and 3PTE (highest correlation) at cutoff
$R_0 = 7\AA$, and structures 
9RNT (lowest correlation) and 3PTE (highest correlation) at cutoff
$R_0=6\AA$.
Heavy solid line refers to crystallographic data, while thin and dashed 
lines refers to GNM approximation.}
\end{figure}

The value of the interaction cutoff for generating the contact matrix 
$\Delta$ has been set to $R_0 = 7\AA$ as 
customarily in such kind of studies. The cutoff choice, which 
affects the overall GNM performance, is generally tested through the
correlation coefficient $\rho$ \cite{Phillips}
\begin{equation}
\rho = \frac{\sum_i(B_i-\langle B \rangle)(X_i-\langle X \rangle)}
      {\sqrt{\sum_{ij}(B_i-\langle B \rangle)^2(X_j-\langle X \rangle)^2}}
\label{eq:rho}
\end{equation}
between experimental ($X_i$) and theoretical ($B_i$) B-factors. 
The sum runs of over the number of protein residues, and  
and $\langle X \rangle$, $\langle B \rangle$ indicate the average
values.
Our data set contains only those protein structures 
with a coefficient $\rho$ greater than 0.5 
(see last column of Tables~\ref{tab7} and \ref{tab6}.)
this should, in principle, ensure that GNM correctly reproduces
C$_{\alpha}$ fluctuations for each selected protein. However 
since we shall study two different cutoffs, we decided to include 
even those few structures, such as 9RNT, 1A47, and 1CDG, that have a 
$\rho>0.5$ for one cutoff and $\rho<0.5$ for the other one.
The few instances of the agreement between B-factors from GNM and 
crystallography are shown in figure~\ref{fig:bfact}, where we display 
the best and the worst cases with respect to the coefficient $\rho$.
 
For each protein, the optimal value of the spring constant $\gamma$
was obtained through a least-square fitting to the experimental B-factors
expressed by formula
\begin{equation}
\frac{k_B T}{\gamma} = \frac{1}{8 \pi^2} \frac{\sum_i B_i X_i}
                        {\sum_i B_i^2}
\end{equation}
The values of $k_BT/\gamma$, besides being an essential ingredient for
the real application of GNM method, are also an indication of the protein
global flexibility and
allows for a direct comparison among all the considered structure.

\begin{figure}
\includegraphics[clip=true,width=0.6\columnwidth,keepaspectratio]
{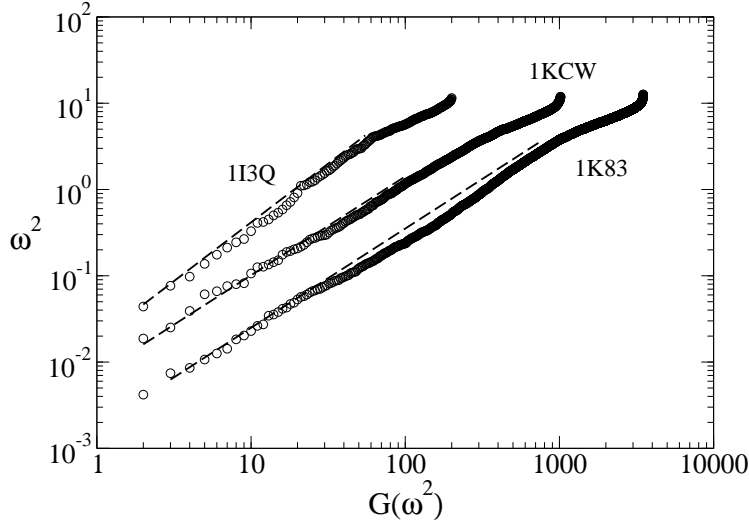}
\caption{
\label{fig:spect}
Log-log plot of GNM-harmonic spectrum referred to three proteins with
different sizes, 1IQQ (N=200), 1KCW (N=1017) and 1K83 (N=3494).
On vertical axis, we report the cumulated distribution $G(\omega)$
of vibrational modes. Low frequencies
regions clearly exhibit a power-law behaviour, and  
dashed lines indicates the best-fits of the power-law   
whose exponent is the spectral dimension.}
\end{figure}

The spectral dimension $\bar{d}$ was estimated via a power-law fitting of
the low frequency behaviour of the cumulated
density of modes $G(\omega)$, namely the integral of $g(\omega)$.
Indeed, due to relation~(\ref{eq:dos}),
$G(\omega) \sim \omega^{\bar{d}}$ at small
arguments (see Fig.~\ref{fig:spect}). 
The harmonic spectrum, obtained within
the GNM, for three proteins with sizes, small, medium and large, 
respectively is plotted in figure~\ref{fig:spect}, where  
low-frequency regions clearly exhibit the power law behavior whose 
exponent is the spectral dimension $\bar {d}$.
\section{RESULTS AND DISCUSSION}

Our statistical analysis for the whole dataset of proteins and  
cutoffs $R_0 = 7\AA$ is summarized in Table~\ref{tab7}, where
we report the spectral
dimension and its corresponding error, the estimate for $k_BT/\gamma$, and
finally the correlation coefficient. 
To test the robustness of our results, we have repeated the same 
analysis at a slightly different cutoff,
namely at $R_0 = 6\AA$, which yields a smaller but still good
correlation between experimental and theoretical B-factors (Tab.~\ref{tab6}). 
Errors on $\bar{d}$-values, in both tables, were estimated to cover
the uncertainty due to the choice of the fitting region for the
power-law, because the slope of the linear-fitting 
(see Fig.~\ref{fig:spect}) can change even sensibly upon varying this region.
Furthermore, error bars take into account also correlation data ($\rho$) 
which indicates how GNM can faithfully reproduce the low-energy
deformations of a given protein structure.

The relationship~(\ref{PL}) establishes a rather strong constraint between 
the spectral dimension and the maximum size $N_{max}$ of a protein can
afford.
Since, the stability is supposed to fail when the
fluctuation $\langle r^2 \rangle^{1/2}$ becomes of the same order of
magnitude of the mean distance between non consecutive amino acids
(about $7$~\AA), one can assume that 
\begin{equation}
\frac{2}{\bar{d}}  = 1 + \frac{b}{\ln(N_{max})}. 
\label{div}
\end{equation}
The proportionality constant $b$ depends on the mean 
amino acid spacing, on the spring elastic constant $\gamma$ and temperature 
$T$. However, this dependence is expected to be very weak
(i.e. only logarithmic) and this allows for a comparison of different
proteins without the computation of the specific parameters. 
It should be stressed that 
equation (\ref{div}), being based uniquely on thermodynamics stability, 
can be actually regarded as an upper bound prediction only. 

\begin{figure}
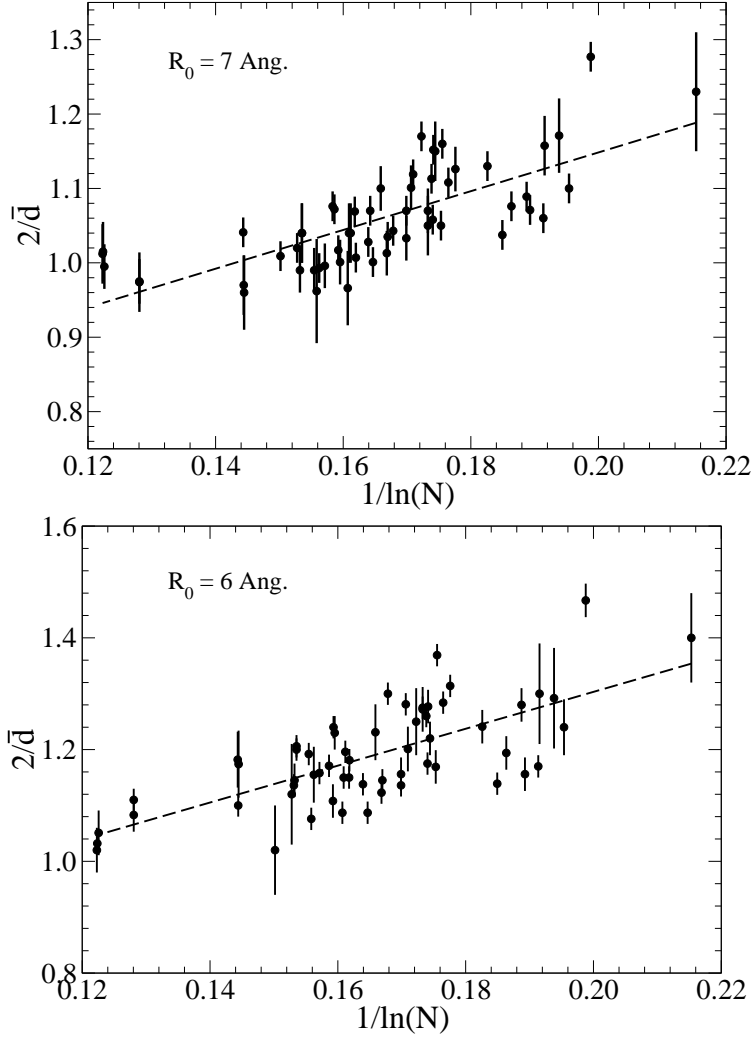
 
\includegraphics[clip=true,width=0.6\columnwidth,keepaspectratio]
{fig3c7.eps}
\includegraphics[clip=true,width=0.6\columnwidth,keepaspectratio]
{fig3c6.eps}
\label{fig:d_of_N}
\caption{Linear plot showing the dependence of the spectral dimension on
protein size. The dashed line, indicating the behaviour~\ref{parb}, is a 
best fit with a correlation coefficient $0.73$ and $0.72$ for cutoffs 
$7\AA$ and $6\AA$ respectively.}
\end{figure}

Figure~\ref{fig:d_of_N} verifies the prediction drawn form the 
thermodynamical stability argument and shows the final result of our 
analysis. 
We plot the quantity $2/d$ versus $1/\ln(N)$ as suggested by relation 
(\ref{div}): indeed, if Eq.~(\ref{div}) holds, we should obtain a straight 
line crossing the y-axis at $1$ for zero abscissa.
As matter of fact, our data are well fitted by a straight line,  
but, with an offset with respect to the equation~(\ref{div}) 
\begin{equation}
\frac{2}{\bar{d}} = a + \frac{b}{\ln(N)}.  
\label{parb}
\end{equation}
For case of a cutoff $R_0= 7\AA$, best-fit values of the parameters are 
$a=0.63$, $b = 2.61$, with a  correlation coefficient $0.73$.  
For the cutoff $R_0= 6\AA$, we obtained the values $a=0.63$, $b = 3.40$.
with a correlation $0.72$.
Interestingly, the linear behaviour predicted by Eq.~(\ref{div})
is confirmed for two different cutoffs with a correlation larger than 
$0.7$, providing a strong evidence of the robustness of the result.

\section{Conclusions}
We applied the Gaussian Network Model (GNM) to investigate the influence of 
native state topology on thermodynamical stability for a set of folded 
proteins with a very different sizes, ranging from 100 to 3600. 
Employing the GNM is appropriate in this type of study because 
such a model correctly accounts for the topological 
features of the native protein conformations. 
Our results show that the spectral dimension $\bar{d}$, which is
sensitive to the large scale topology of a geometrical structure, is one  
parameter governing the low-energy fluctuations of a given protein 
structure.   
As a consequence, one can derive an instability criterion for proteins, 
based only on topological considerations, which is the analogous of 
Peierls' criterion developed for ordered crystalline structures. 
The criterion easily predicts a non-trivial logarithmic dependence 
of the spectral dimension on the length of a protein.   
This further confirms the lack of universality for the spectral dimension 
of proteins \cite{Avraham}, an issues already addressed in previous 
studies \cite{Haliloglu}. 
We verified that such a logarithmic dependence is really observed, 
within statistical and systematic errors, for the whole set of selected 
proteins. Furthermore, the dependence is robust because it applies even with
alteration of the 
interaction cutoff which is the most critical parameter to the GNM 
applicability. We can conclude that 
the relation between spectral dimension and length of proteins is not 
an artifact due to a particular cutoff choice, providing that a 
significant correlation is maintained between experimental and theoretical 
B-factors. We verified that, at a larger cutoff, the scaling 
behaviour~(\ref{parb}) is preserved, although the spectral dimension grows 
due to the increase in the average connectivity of the elastic network.

The result expressed by Eq.~(\ref{parb}) deserves some comments.

Equation~(\ref{parb}) is in agreement with
the upper bound represented by Eq.~(\ref{div}), supporting the relevance of 
topological thermal
instability as a constraint to protein geometry. More importantly, not only is the
upper bound satisfied, but the experimental points lie on a straight line
parallel to the upper bound line of Eq.~(\ref{div}). 
This suggests a more fundamental role
of topological stability: the protein tends to arrange
topologically in such a way to reach the minimum value compatible with 
stability constraints. 
In other words, for any fixed length, it tends to the most swollen
state which remains stable with respect to thermal fluctuations.
  
An interesting point is the meaning of the offset $a-1$ which would be $0$ 
according to Eq.~(\ref{div}). Its
positive value could have different explanations, but its universal nature (it
is a ``protein-independent'' because is a global shift) must be due to a very 
general mechanism. A rather obvious reason is the contribution of anharmonic
interactions at finite temperatures; a more intriguing one could be an
effective longer range interaction due to the presence of bound water 
molecules around the external amino acids, which could change the effective 
form of the interaction matrix $\Lambda$. This hypothesis is also suggested 
by the physical interpretation of $b$ as an anomalous dimension exponent,
typically related to a renormalized interactions~\cite{Golden}.
However, the most intriguing evidence relies on the regression coefficient
Independently of the physical origin of $b$, its high value strongly supports
the existence of a thermodynamic stability threshold, dependent on the 
topology of the folded state, for the size of proteins.

%

\begin{table} 
\includegraphics[clip=true,width=0.4\columnwidth,keepaspectratio]
{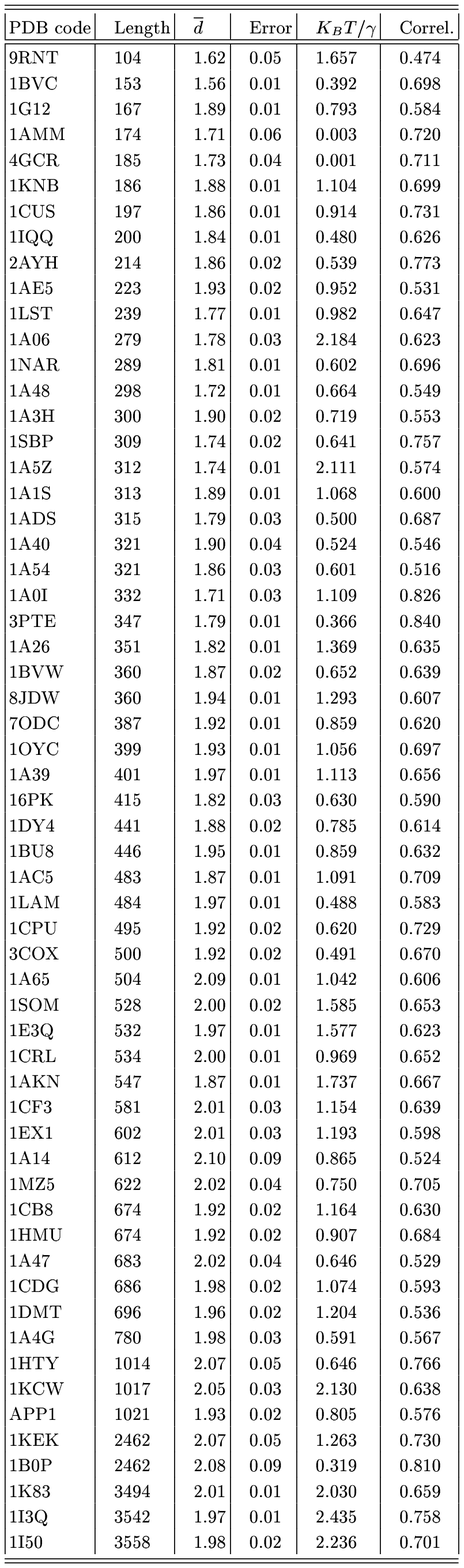}
\caption{\label{tab7}
List of processed native protein structures from Brookheaven PDB,
with their length, the corresponding spectral dimension estimated by GNM
approach with cutoff $R_0 = 7\AA$, error on its determination, 
parameter $k_BT/\gamma$ and correlation $\rho$ ~(\ref{eq:rho}).}
\end{table}

\begin{table} 
\includegraphics[clip=true,width=0.4\columnwidth,keepaspectratio]
{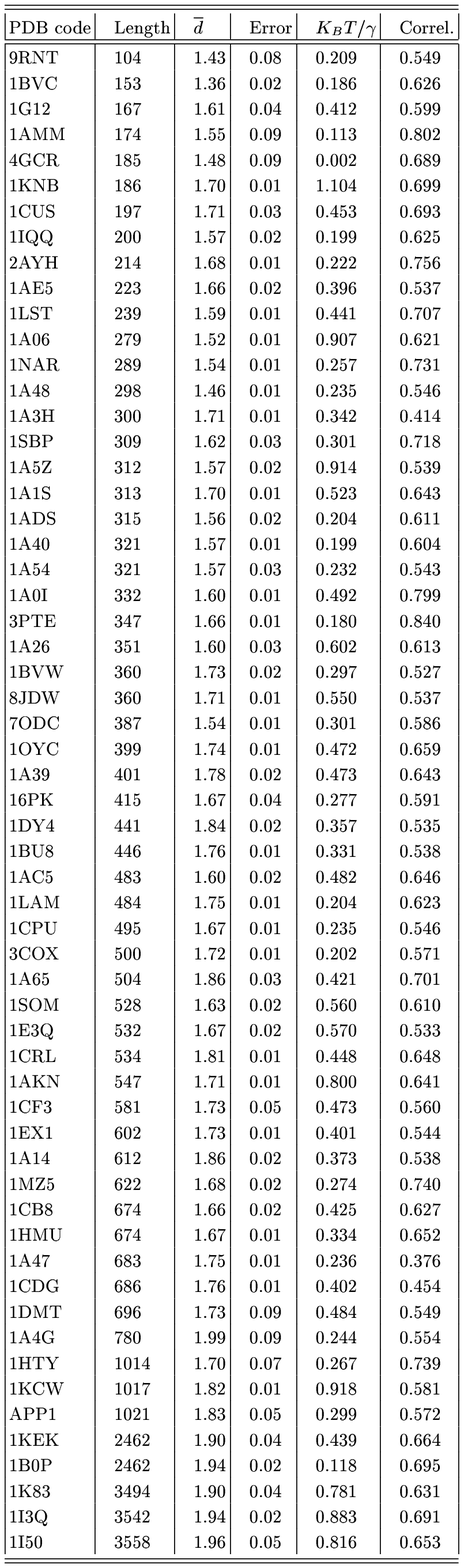}
\caption{\label{tab6}
List of processed native protein structures from Brookheaven PDB,
with their length, the corresponding spectral dimension estimated by GNM
approach with cutoff $R_0 = 6\AA$, error on its determination, 
parameter $k_BT/\gamma$ and correlation $\rho$ ~(\ref{eq:rho}).}
\end{table}
\end{document}